\documentstyle[prd,aps]{revtex}
\input psfig
\pssilent
\begin{document}

\title{Chern--Simons states in spin-network quantum gravity}

\author{Rodolfo Gambini, Jorge Griego
\\{\em Instituto de F\'{\i}sica, Facultad de
Ciencias,\\Tristan Narvaja 1674, Montevideo, Uruguay}}

\author{Jorge Pullin\\  {\em Center for Gravitational Physics and Geometry}\\
{\em Department of Physics, 104 Davey Lab,}\\ {\em The Pennsylvania State 
University,}\\
{\em University Park, PA 16802}}

\maketitle
\vspace{-6cm} 
\begin{flushright}
\baselineskip=15pt
CGPG-97/3-4  \\
gr-qc/9703042\\
\end{flushright}
\vspace{5cm}
\begin{abstract}
In the context of canonical quantum gravity in terms of Ashtekar's new
variables, it is known that there exists a state that is annihilated
by all the quantum constraints and that is given by the exponential of
the Chern--Simons form constructed with the Asthekar connection. We
make a first exploration of the transform of this state into the
spin-network representation of quantum gravity. The discussion is
limited to trivalent nets with planar intersections. We adapt an
invariant of tangles to construct the transform and study the action
of the Hamiltonian constraint on it. We show that the first two
coefficients of the expansion of the invariant in terms of the inverse
cosmological constant are annihilated by the Hamiltonian
constraint. We also discuss issues of framing that arise in the
construction.
\end{abstract}

\vspace{1cm}

It was noted some time ago \cite{BrGaPunpb} that the exponential of
the Chern--Simons form
\begin{equation}
\Psi^{CS} [A] = \exp\left(-{6 \over \Lambda } {\rm Tr} 
[\int A\wedge d A + {2 \over 3} A \wedge A \wedge A]\right)
\end{equation}
was annihilated by all the constraints of
canonical quantum gravity (with a cosmological constant) in the
connection representation constructed via the Ashtekar \cite{As86}
formulation of canonical general relativity. Starting with that
observation there have been several attempts
\cite{BrGaPunpb,BrGaPuessay,DiBaGrGaPu,DiBaGrGa} to understand the
counterpart of this state in the loop representation.

In order to compute the counterpart one has to consider the loop 
transform of such a state,
\begin{equation}
\Psi(\gamma) = \int DA \exp\left(-{6 \over \Lambda }{\rm Tr}[\int A\wedge d 
A + {2 \over 3}
A \wedge A \wedge A]\right) W_\gamma[A].
\end{equation}
where $W_\gamma[A]$ is the Wilson loop, ie the trace of the holonomy of 
the connection along the loop $\gamma$. 

Wilson loops are constrained by a set of relations known as the
Mandelstam identities. The presence of these identities implies that
not every function of a loop $\gamma$ is an admissible wavefunction in
the loop representation.  It has been shown that a set of gauge
invariant quantities that is free of Mandelstam identities can be
constructed from loops through the use of the spin-network basis
\cite{RoSmspin}. In this context spin networks consist of lines
connected at vertices living in three dimensional space, with each
line having associated a holonomy in a given representation of
$SU(2)$. One can then construct gauge invariant quantities by
contracting the holonomies with appropriate intertwiners provided by
the theory of angular momentum recoupling at the intersections. We
will call the resulting invariant a ``Wilson-net'' and denote it
$W_\Gamma[A]$ where $\Gamma$ is a spin network.

One can then introduce a new representation for quantum gravity in which
wavefunctions are labelled by spin networks via a transform similar to the
one introduced above. In particular, for the Chern--Simons state,
\begin{equation}
\Psi(\Gamma) = \int DA \exp\left(-{6 \over \Lambda }
{\rm Tr}[\int A\wedge d A + {2 \over 3}
A \wedge A \wedge A]\right) W_\Gamma[A].
\end{equation}

With the introduction of rigorous measures in the space of connections 
modulo gauge transformations \cite{AsLe}, hopes were raised that the 
above functional integral could be computed in a rigorous way. It turns
out however, that because the Chern--Simons form is only gauge invariant
when integrated over the manifold, this makes use of the rigorous measure
theory a nontrivial task that has not been accomplished up to present.

On the other hand, the above integral (in the context of ordinary
loops) has been explored by a variety of methods, which can be
summarized in three groups. One of them considers a two dimensional
slice of the Chern--Simons theory and studies the monodromies that
result from exchanges of the punctures that the Wilson loops form with
the slice \cite{Wi}. The integral can also be evaluated perturbatively
using Feynman diagrammatics \cite{CoGuMaMi,BaNa}. This produces a
result that is an explicit function given by loop integrals of certain
kernels. The last technique (variational calculus) consists of
studying infinitesimal deformations of the loop that appears in the
integral and computing relations (skein relations) between the values
of the wavefunction when one substitutes over and under-crossings in
the loop by intersections \cite{Sm,GuMaMi,BrGaPunpb}. In some cases the
infinitesimal results can be exponentiated to yield finite results
\cite{GaPucmp}. On the other hand it was shown in \cite{GaPucmp} that
variational techniques offer a certain amount of regularization freedom
in the definition of the transform. Therefore the possibility exists that
there are really various possible ``transforms'' of the Chern--Simons
state, essentially corresponding to the introduction of different measures
in the computation of the functional integration.

In the context of spin networks, the transform of the Chern--Simons state
has been explored by Kauffman and Lins \cite{KaLi} through the use of 
tangle-theoretic techniques. What we will do in this paper is to reinterpret
the tangle results in terms of spin networks and 
study if the resulting invariant is annihilated by the Hamiltonian
constraint of general relativity in the context of spin networks
originally proposed by Thiemann \cite{Th}.

To start setting up the structure we need, we will begin by pointing
out certain conventions. To define an invariant of a trivalent spin
network, one needs a $3-j$ symbol which is the intertwiner at the
intersections, and also an assignment of an ``orientation'' at the
vertices, to determine which incoming holonomy corresponds with each
entry of the symbol (there are two cyclic possibilities). The
choice we make is to order the three entries by considering the
clockwise ordering of the three incoming lines. If one considers the
holonomies connected with intertwiners the resulting object will in 
general be orientation-dependent. For the $SU(2)$ case it can be seen
that the orientation dependence boil down to an overall sign of the 
invariant. We will define our invariants with an additional overall
sign in such a way that they are not orientation dependent. 
The overall factor will be chosen in such a way that if one considers
the invariant when the connection is flat its value corresponds to 
the chromatic evaluation of the graph (see \cite{KaLi}; also
\cite{DePiRo} for some explicit examples).

\begin{figure}
\hspace{3in}\psfig{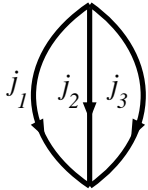}
\caption{An example of spin network.}
\label{spinfig}
\end{figure}

Let us show an example of how to construct a
gauge invariant function associated with a spin network. We will call
these invariant ``Wilson network'' since it naturally generalizes the
idea of Wilson loops to this context. Let us consider a spin network
like the one shown in figure \ref{spinfig}
the resulting Wilson network is given by,
\begin{equation}
W_\Gamma=f(\Gamma) \sum_{m,m'}
\left\{\begin{array}{ccc}
j_1 &j_2&j_3\\
m_1&m_2&m_3\\
\end{array}\right\}
\left\{\begin{array}{ccc}
j_1 & j_2 & j_3\\
-{m'}_1&-{m'}_2&-{m'}_3\\
\end{array}\right\}
H^{j_1}_{m_1\,{m'}_1}
H^{j_2}_{m_2\,{m'}_3}
H^{j_3}_{m_3\,{m'}_3}\label{Wilsonet}
\end{equation}
where  $f(\Gamma)$ the factor we discussed above. For all the networks
we will consider in this paper, the above factor is given (up to a
factor of modulus unity) by the following formula,
\begin{equation}
f(\Gamma)= \pm \prod_{v_i \in {\rm vertices}} 
\sqrt{\Theta_0(v_i) }
\end{equation}
where 
\begin{equation}
\Theta_0(v_i)={(-1)^{j_1+j_2+j_3} (j_1+j_2+j_3+1)! (j_1+j_2-j_3)!
(j_1+j_3-j_2)! (j_2+j_3-j_1)!\over (2 j_1)! (2 j_2)! (2 j_3)!}
\end{equation}
and $j_{1-3}$ are the spins of the strands associated with the vertex
in question.

Ordinary non-intersecting loops are considered in this approach as spin
networks composed by two strands joined at ``double vertices''. If one
evaluates (\ref{Wilsonet}) for that case, one notices that the result
is $(-1)^{2j}$ times the ordinary Wilson loop. 

The normalization we have chosen for the spin networks coincides with
that of De Pietri and Rovelli \cite{DePiRo}. The notation is different
from that of Thiemann \cite{Th} in the sense that he introduces double
vertices along each strand. We only consider double vertices as the
ones we introduced to describe the ordinary Wilson loop, that is, they
always appear in pairs.

The invariant we will consider is defined by the following relations:

$\bullet$  Basic symmetries: 

If one considers the un-knot in the trivial 
representation, the Wilson net is equal to one, and the result of the
transform is chosen to be 
\begin{equation}
E({\rm unknot}^0,k) = 1,
\end{equation}
this being a normalization condition. 

The Wilson nets for $SU(2)$ are real, so $W_\Gamma = \bar{W}_\Gamma$,
and therefore,
\begin{equation}
\bar{E}(\Gamma,k) = E(\Gamma,-k).
\end{equation}

Since the Chern--Simons form changes sign under a parity
transformation,
\begin{equation}
E(\tilde{\Gamma},k) = E(\Gamma,-k),
\end{equation}
where $\tilde{\Gamma}$ is the mirror-image of the spin-net $\Gamma$.

Finally,
\begin{equation}
E(\Gamma,\infty) = W_\Gamma[A=0]=\hbox{chromatic evaluation}(\Gamma).
\end{equation}
This result can be understood remembering that in the case of single
loops in the fundamental representation the parameter $k$ is related
with the parameter of the Kauffman bracket $q$ via $q=e^{-{4i\pi\over
k}}$, so for $k\rightarrow\infty$, $q\rightarrow 1$ and the invariant
is equal to 2, which corresponds (via Giles' \cite{Gi} reconstruction
theorem) to evaluating the Wilson loop for a flat connection. This
translates in the case of spin-nets the chromatic evaluation.

$\bullet$ Factorization.

If one considers two spin-nets that are disjoint (by this meaning
their planar projections lie in two separate half-planes) then,
\begin{equation}
E(\Gamma_1 \cup \Gamma_2) = <W_{\Gamma_1} W_{\Gamma_2}> =
<W_{\Gamma_1} > <W_{\Gamma_2} > = E(\Gamma_1) E(\Gamma_2).
\end{equation}
This was proven by Witten \cite{Wi} and can be intuitively seen by
remembering that since $E$ is diffeomorphism invariant, one could
consider the two spin networks to be as far removed from each other as
possible and then all the ``interactions'' of the Chern--Simons theory
would disappear.

$\bullet$ Twists and crossings.

\begin{figure}[h]
\hspace{3cm}\psfig{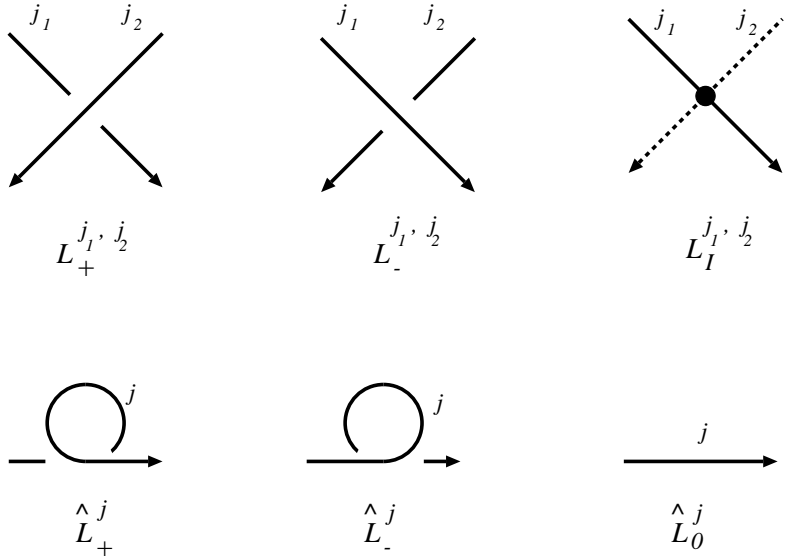}
\caption{The different crossings involved in the skein relations for 
the invariant} 
\label{twists}
\end{figure}

The values of the invariant when  one replaces and under-crossing by and
over-crossing are related through equations that are generalizations of the
usual ``skein relations'' that are satisfied by the Kauffman bracket 
(which is
the transform of the Chern--Simons state in terms of ordinary loops.)
These relations are,
\begin{equation}
E(\hat{L}_\pm^{(j)},k) = \exp\left( \mp {4 i \pi\over k} Q_j\right)
E(\hat{L}_0^{(j)}) = q^{\pm Q_j} E(\hat{L}_0^j)
\end{equation}
where $\hat{L}\pm,\hat{L}_0$ are depicted in figure \ref{twists} and
$\Lambda = 24 i \pi/k$.

For the case of upper and under-crossings, 
\begin{equation}
E(L^{j_1,j_2}_\pm,k) = \sum_{j=|j_1-j_2|}^{j_1+j_2} q^{\mp {1\over 2}
[Q_{j_1}+Q_{j_2}-Q_{j}]} (-1)^{j_1+j_2-j} {\Delta_j
\over \Theta(j_1,j_2,j)} (-1)^{n_1 j_1+n_2 j_2}  E(L_{H}^{j_1,j_2,j},k)
\label{spinskein}
\end{equation}
\begin{figure}[h]
\hspace{3in}\psfig{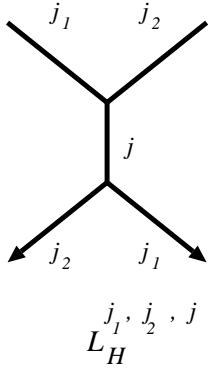}
\caption{The double-trivalent element that arises in the computation of the
upper and under-crossings.}
\label{hache}
\end{figure}
\noindent 
where $n_1$ and $n_2$ are the number of two-valent vertices that have
been
eliminated by the introduction of the double-trivalent vertex (essentially
this means that if one had a single loop with a crossing, it would
have had two double-valent crossings, and when the crossing is
replaced with the $L_H$ they would disappear, giving rise to an
overall factor due to normalization; the situation could also arise
from two separate loops that crossed). In the above expression $\Theta$ and 
$\Delta_j$ are the evaluations of the invariant for the theta knot and the
un-knot,
\begin{eqnarray}
\Delta_j &=&E\left(\raisebox{-5mm}{\psfig{figure=unknotj.eps,height=10mm}}
\,,\,k\right) \\
\Theta(j_1,j_2,j) &=& E\left(\raisebox{-12.5mm}
{\psfig{figure=thetajj1j2.eps,height=25mm}}\,,\,k\right)
\end{eqnarray}
and we will detail later how to evaluate the invariant for the diagrams
shown.

One can obtain these relations using variational techniques \cite{GaPucmp}
and using a generalization of the Fierz identity for higher representations
of $SU(2)$,
\begin{equation}
\tau^i_{(j_1)\,A}{}^B \tau^i_{(j_2)\,C}{}^D = 
{1 \over 2} \left\{ 
\sum_{J=|j_1-j_2|}^{j_1+j_2} Q_J \sum_{M=-J}^J 
<j_1\,A\,j_2\,C | J\,M>
<J\,M|j_1\,B\,j_2\,D> - (Q_{j_1}+Q_{j_2}))
 \delta_A^B \delta_C^D \right\}.
\end{equation}

$\bullet$ Recoupling identities.

The additional identities that will define the invariant cannot be obtained
in a simple explicit way using variational calculations. We just impose them
at the level of spin nets and assume they define a regularization in terms
of which one could perform variational calculations for their evaluation.
Together with the above relations they completely characterize the invariant,
ie, they allow us to evaluate it for all possible spin networks in three
dimensions.

The first relation allows to eliminate a loop from a line,
\begin{equation}
E\left(\raisebox{-12.5mm}{\psfig{figure=thetaopen.eps,height=25mm}}\,,\,
k\right)
= \delta_{j\,j'} 
{E\left(\raisebox{-12.5mm}{\psfig{figure=thetajj1j2.eps,height=25mm}},k\right) 
\over 
E\left(\raisebox{-5mm}{\psfig{figure=unknotj.eps,height=10mm}}\,,\,k\right)} 
= \delta_{j\,j'} 
{\Theta(j,j_1,j_2)  \over \Delta_j} 
E\left(\quad
\raisebox{-7mm}{\psfig{figure=linej.eps,height=15mm}}\quad,\,k\right)
\label{untheta}
\end{equation}
and this relation can be simply understood by ``closing up'' the upper and
lower strand to form the theta and the unknot.

The second relation is,
\begin{equation}
E\left(\raisebox{-10mm}{\psfig{figure=wishbone.eps,height=20mm}}\,,\,
k\right)=\sum_{j={|j_1-j_4|}}^{|j_1+j_4|}
\left\{
\begin{array}{ccc}
j_2&j_1&j\\
j_4&j_3&l
\end{array}
\right\}_q 
E\left(\raisebox{-10mm}{\psfig{figure=doubley.eps,height=20mm}}\,,\,k\right)
\label{wishbone}
\end{equation}
where the expression in curly braces is the q-deformed Racah symbol,
which is defined as,
\begin{equation}
\left\{
\begin{array}{ccc}
j_2&j_1&j\\
j_4&j_3&l
\end{array}
\right\}_q 
= {E\left(
\raisebox{-10mm}{\psfig{figure=tet.eps,height=20mm}}\,,\,k\right) 
E\left(
\raisebox{-5mm}{\psfig{figure=unknotj.eps,height=10mm}}\,,\,k\right) \over
E\left(
\raisebox{-10mm}{\psfig{figure=thetajj1j4.eps,height=20mm}}\,,\,k\right) 
E\left(
\raisebox{-10mm}{\psfig{figure=thetajj2j3.eps,height=20mm}}\,,\,k\right) }.
\end{equation}

The left diagram in the numerator is known as the tetrahedron and denoted,
\begin{equation}
{\rm Tet}\left(
\begin{array}{ccc}
j&j_4&l\\
j_2&j_3&j
\end{array}\right) = 
E\left(
\raisebox{-10mm}{\psfig{figure=tet.eps,height=20mm}}\,,\,k\right) 
\end{equation}
and an explicit evaluation for it is given in the book by Kauffman and
Lins \cite{KaLi}, paragraph 8.5.

The above relations completely characterize the invariant. In
particular, we can now evaluate its value for the unknot of $j$
valence, $\Delta_j$.  In order to do that one considers $j$ parallel
loops of valence $1/2$ and uses equation (\ref{wishbone}) with $k=0$
to convert two parallel lines of valence $1/2$ to a line of valence
$1$. The resulting object (for two loops) is a theta diagram with
$j=1$, which we can transform using (\ref{untheta}) to a single loop
of higher valence. The reason why we emphasize this construction is to
notice that for specifying a loop (or for that matter a single strand)
of higher valence one requires to specify a family of ``parallel''
loops that do not intertwine (if they did, the value assigned to the
loop of higher valence would be different).  This indicates that the
invariant we are constructed is in a very basic nature an invariant of
framed loops. One needs to prescribe a framing for each higher valence
line, even if no twists of it are present. Of course the invariant is
an invariant of framed loops in the ordinary sense, since the addition
of a twist changes its value. But there is this deeper, more basic
framing associated with a single line. This is a well known effect, it
is just a fact that the invariant we are computing is really an
invariant of $q$-deformed spin networks \cite{MaSm} which require such
framings. The computation of $\Theta$ is also straightforward, we
refer the reader to chapter 6.3 of Kauffman and Lins \cite{KaLi}.

This weakens considerably the case for studying these invariants in
the context of ordinary quantum gravity where the objects in question
are non-q-deformed spin networks. However, Major and Smolin
\cite{MaSm} have proposed that q-deformed spin networks might play a
role in quantum gravity. If this were the case, the invariant
considered could be a candidate to quantum state. Based in part on
this motivation we will explore the action of Thiemann's \cite{Th}
Hamiltonian constraint on this state. Of course, Thiemann's
Hamiltonian does not necessarily a priori know how to act on framed
spin networks. The approach we will take is to assume that the action
is the same as on ordinary spin networks. Clearly one could define a
Hamiltonian in the context of framed spin networks that would act in
such a way.

The explicit action of Thiemann's Hamiltonian in terms of spin networks
has been worked out by Borissov, De Pietri and Rovelli \cite{BoDiRo}. 
The operator only acts at vertices by adding a line forming a triangle
with the vertex in the three possible orientations and multiplying times
a factor,
\begin{eqnarray}
\hat{H} \quad\psi\left(
\raisebox{-7mm}{\psfig{figure=trival.eps,height=20mm}}\right)  =
{2 i l_0 \over 3}
\sum_{s=\pm 1}
\sum_{t=\pm 1}
&&\left[
A(q,t,r,s,p) 
\quad\psi\left(
\raisebox{-7mm}{\psfig{figure=triang1.eps,height=20mm}}\right)  
+A(p,t,q,s,r)
\quad\psi\left(
\raisebox{-7mm}{\psfig{figure=triang2.eps,height=20mm}}\right) \right.
\nonumber\\
&&\quad\left.
+A(r,t,p,s,q)
\quad\psi\left(
\raisebox{-7mm}{\psfig{figure=triang3.eps,height=20mm}}\right)  
\right]
\end{eqnarray}
where the coefficients $A$ are explicitly given in \cite{BoDiRo}. This
is the action of the Euclidean part of Thiemann's Hamiltonian constraint.
Since the Chern--Simons state is annihilated by the Euclidean Hamiltonian
constraint in the connection representation, we will restrict our analysis
to the Euclidean part here.

Now, for our invariant $E(\Gamma,k)$, the recoupling relations allow to
connect the value of the invariant for the trivalent vertex with the
extra line with the value for just a plain trivalent vertex, through
the formula,
\begin{equation}
E\left(
\raisebox{-7mm}{\psfig{figure=triang1.eps,height=20mm}}\,,\,k\right) = 
{{\rm Tet}\left(
\begin{array}{ccc}
q&q+t&1\\
r&r+s&p
\end{array}\right)\over \Theta(p,q,r)}
E\left(
\raisebox{-7mm}{\psfig{figure=trival.eps,height=20mm}}\,,\,k\right),  
\end{equation}
let us denote the prefactor in this equation as $\beta(p,q,r,k)$ (it
is a function of k).

If one now combines this expression with the action of the
Hamiltonian, one basically proves that the action of the Hamiltonian
reduces to an overall multiplication times a coefficient that is a
linear combination of tetrahedra with the coefficients $A$. One can
examine this expression order by order in the coupling constant $k$,
in effect examining the action of the Hamiltonian constraint on the
various powers of the expansion of the invariant $E$ in terms of
$k$. For order $k^0$ the invariant reduces to the chromatic evaluation
of the knot. Such an invariant can be thought of as the counterpart in
the spin network representation of a distributional state in the
connection representation given by $\psi[A] = \delta[A-{\rm flat}]$.
That is a state ``peaked'' around flat connections. Such a state is
obviously a solution of the Hamiltonian constraint in the factor
ordering in which the curvature appears to the right (this is the 
ordering that Thiemann chose for his Hamiltonian). It can actually
be checked by explicit computation that the linear combination of 
$A's$ with the zeroth order expansion of the $\beta$ prefactor
vanishes \cite{Di}. This is therefore a solution of the Euclidean
Hamiltonian constraint. In fact, the zeroth order coefficient (chromatic
evaluation) is framing independent so this is a genuine state of 
quantum gravity in the spin-network representation.

What happens to higher orders? The computations are quite involved.
However, we can immediately reach a conclusion for the first order in
$k$. Because the $\beta$ coefficient is chiral (a knot and its mirror
image give the same result for it) and a chiral operation on the
invariant (as we discusses above) is tantamount to a change of sign in
$k$, it is the fact that the $\beta$ prefactor should have zero first
coefficient in its expansion in terms of $k$. Therefore the action of
the Hamiltonian constraint on $E(\Gamma,k)$ to first order in $k$
reduces to its zeroth order action, and is again zero. From this we
can conclude that the first coefficient of the expansion of the
invariant $E(\Gamma,k)$ in terms of $k$ is also annihilated by the
Hamiltonian constraint.  Here one should exercise some care since this
invariant is not framing independent and therefore is not a genuine
state in terms of ordinary spin networks. The only conclusion one can
reach is that if one were to consider a framed spin network
representation of quantum gravity and define Thiemann's Hamiltonian
constraint in a manner analogous to the non-framed case, this state
would be a candidate to a solution.

In summary, we have constructed in a rather ad-hoc way an invariant
that is a candidate to be the transform of the Chern--Simons state in
terms of spin network. The ad-hoc portions of the construction can be
ascribed to choices of regularization in the definition of the path
integral. The choices made allow the invariant to have the properties
of recoupling theory, which in turn reduces the action of the
Hamiltonian constraint to a multiplicative operator. As a consequence
we are able to find solutions of the Hamiltonian constraint. The price
to pay for the choices made in the definition of the invariant is that
it is really an invariant of framed spin nets and therefore not giving
rise to genuine states (with the exception of the zeroth coefficient)
of ordinary (non q-deformed) quantum gravity. This raises the question
of if other choices of regularization could not allow the definition
of an invariant that is a genuine invariant of spin nets. This issue
is currently under consideration. In principle the systematic use of
the variational techniques of \cite{GaPucmp} could in principle allow
the construction of such an invariant. It is quite unlikely, however,
that the action of the Hamiltonian for such an invariant will reduce
to just a multiplication times a factor. On the other hand, the
resulting invariant could potentially be made framing independent by
the removal of the  twist dependences much in the same spirit as the
Jones polynomial is obtained by removing the twist dependence of the
Kauffman bracket in the context of ordinary loops. Such a result would
be of interest in its own right, even if it fails to produce solutions
of the Hamiltonian constraint of quantum gravity.

We wish to thank Roumen Borissov,
Roberto De Pietri and Carlo Rovelli for making their work available
prior to publication and for discussions.  We are also grateful to
Roberto De Pietri for checking for all valences the zeroth order
cancellation of the action of the Hamiltonian.  This work was
supported in part by grants NSF-INT-9406269, NSF-PHY-9423950,
NSF-PHY-9396246, research funds of the Pennsylvania State University,
the Eberly Family research fund at PSU and PSU's Office for Minority
Faculty development. JP acknowledges support of the Alfred P. Sloan
foundation through a fellowship. We acknowledge support of Conicyt and
PEDECIBA (Uruguay).

\end{document}